\documentclass{article}
\usepackage{spconf,amsmath,graphicx,hyperref}
\usepackage{cite}
\usepackage{amssymb,amsfonts}
\usepackage{algorithmic}
\usepackage{textcomp}
\usepackage{xcolor}
\usepackage{multirow}
\usepackage{pifont}
\usepackage{makecell}
\usepackage{subcaption}
\usepackage{threeparttable}
\usepackage{booktabs}
\usepackage{url}
\usepackage[table]{xcolor}
\usepackage[ruled,vlined]{algorithm2e}
\newcommand{\cmark}{\ding{51}}
\newcommand{\xmark}{\ding{55}}
\title{Train Short, Infer Long: Speech-LLM Enables Zero-Shot \\ Streamable Joint ASR and Diarization on Long Audio}
\name{Mohan Shi$^{1*}$, Xiong Xiao$^2$, Ruchao Fan$^2$, Shaoshi Ling$^2$, Jinyu Li$^2$\thanks{$^*$Work done during an internship at Microsoft.}}
\address{$^1$University of California, Los Angeles, USA \\ $^2$Microsoft CoreAI, Redmond, USA}
\begin{document}
\ninept
\maketitle
\begin{abstract}
Joint automatic speech recognition (ASR) and speaker diarization aim to answer the question ``who spoke what'' in multi-speaker scenarios. In this paper, we present an end-to-end speech large language model (Speech-LLM) for \textbf{J}oint str\textbf{E}amable \textbf{DI}arization and a\textbf{S}r (JEDIS-LLM). The model is trained only on short audio under 20s but is capable of streamable inference on long-form audio without additional training. This is achieved by introducing a Speaker Prompt Cache (SPC) with an on-the-fly update mechanism during chunk-wise streaming inference, inspired by the autoregressive nature of LLMs. 
% Furthermore, by replacing the Speaker Prompt Cache with manually segmented audio clips and their corresponding transcriptions, the model can seamlessly integrate speaker profiles without retraining.
The SPC also allows the seamless use of pre-enrolled speaker profiles which is common in many scenarios like meeting transcription.
To further enhance diarization capability, we incorporate word-level speaker supervision into the speech encoder during training.

Experimental results demonstrate that our system outperforms strong baselines, including Sortformer and Meta-Cat in the local setting on audio up to 20s,
% short audio in the offline setting,
and DiarizationLM on long-form audio, despite being fully end-to-end and streamable while DiarizationLM follows a cascaded offline pipeline. To the best of our knowledge, this is the first work enabling zero-shot streamable joint ASR and diarization on long audio using a Speech-LLM trained only on short audio, achieving state-of-the-art performance.
% and representing a significant breakthrough.

% In this paper, we introduce Phi-4-SD, an end-to-end speech large language model (Speech-LLM) that unifies streaming and non-streaming joint ASR and diarization. The model is built upon the prevailing Speech-LLM architecture and is fine-tuned on speaker-attributed transcriptions using low-rank adaptation (LoRA), while jointly training the speech encoder and modality projector. To enhance diarization without severely degrading ASR, we incorporate word-level speaker supervision into the speech encoder.
% In addition, we propose a Speaker Prompt Cache with an update mechanism that leverages the autoregressive nature of LLMs, enabling chunk-wise streaming inference on long-form audio and seamless speaker profile integration without additional training. Experimental results show that our model outperforms strong baselines, including Sortformer and Meta-Cat on short audio in the offline setting and DiarizationLM on long-form audio, despite our model being end-to-end streaming while DiarizationLM is cascaded offline. To our knowledge, this is the first unified framework for joint ASR and diarization in both streaming and non-streaming settings with an LLM, achieving state-of-the-art performance and representing a significant breakthrough.
\end{abstract}
\vspace{-0.1cm}
\section{Introduction}
\label{sec:intro}
\vspace{-0.2cm}
With advances in deep learning and automatic speech recognition (ASR)~\cite{li2022recent,GulatiQCPZYHWZW20,YaoGY0KYJLP24}, multi-speaker scenarios such as meetings and conversations have received increasing attention. Speaker diarization~\cite{ParkKDHWN22,FujitaKHNW19,MedennikovKPKKS20} aims to identify ``who spoke when'' in a recording. Combined with ASR, it forms the joint ASR and diarization task~\cite{ShafeySS19,KandaGWMCZY20}, also known as speaker-attributed ASR~\cite{KandaGWMCZY20,KandaYGWMCY21,ShiD0YLZ0023,LiangSYLZDCXQWCLYB23}, which addresses ``who spoke what'' and is crucial for understanding multi-speaker conversations.

% Previous works combined the outputs of independent ASR and diarization systems to obtain speaker-attributed transcriptions~\cite{KandaXWZGWMCY21,YuDZL022,ShiZDYCZD23}. Such cascaded systems are suboptimal due to error propagation. Recently, methods such as Sortformer~\cite{abs-2409-06656} and Meta-Cat~\cite{WangWDPKMHKBG25} have integrated speaker IDs seamlessly into multi-talker transcriptions, enabling direct training for speaker-attributed transcriptions and showing reasonable performance in the local setting with audio up to 20s. However, these approaches still rely on two pre-trained encoders, one for ASR and the other for speaker diarization, and their representations must be fused for decoding. A more critical limitation is that they are not suitable for long-form audio requiring global diarization.

Previous works combined independent ASR and diarization outputs to obtain speaker-attributed transcriptions~\cite{KandaXWZGWMCY21,YuDZL022,ShiZDYCZD23}, but such cascaded systems suffer from error propagation. Recent approaches, such as Sortformer~\cite{abs-2409-06656} and Meta-Cat~\cite{WangWDPKMHKBG25}, integrate speaker IDs directly into multi-talker transcriptions, enabling end-to-end training and achieving reasonable performance on short audio (up to 20~s). However, these methods still rely on two separate pre-trained encoders for ASR and diarization, whose representations are fused during decoding. More importantly, they are not suitable for long-form audio that requires global diarization. Streaming Sortformer~\cite{abs-2507-18446} introduces a cache mechanism for global diarization, but it does not incorporate the ASR task.

% Previous works combined the outputs of independent ASR and diarization systems to obtain speaker-attributed transcriptions~\cite{KandaXWZGWMCY21,YuDZL022,ShiZDYCZD23}. Such cascaded systems are suboptimal due to error propagation. Other approaches explored end-to-end SA-ASR systems~\cite{KandaYGWMCY21,ShiD0YLZ0023}, often employing two branches: one predicts multi-talker transcriptions in serialized output training (SOT)~\cite{KandaGWMY20} style, while the other predicts speaker IDs for each text token. These two branches are jointly optimized. However, these systems still treat transcription and speaker ID prediction as separate outputs and often require pre-extracted speaker embeddings. More recently, works such as Sortformer~\cite{abs-2409-06656} and Meta-Cat~\cite{WangWDPKMHKBG25} have integrated speaker IDs directly into multi-talker transcriptions to obtain speaker-attributed transcriptions and have shown reasonable performance on short audio. However, these methods still rely on two pre-trained encoders, one for ASR and one for speaker diarization, whose representations are fused for decoding, and they are not suitable for long-form audio.

The long-context reasoning and cross-utterance awareness abilities of large language models (LLMs)~\cite{achiam2023gpt,dubey2024llama,abdin2024phi} make them well-suited for multi-talker conversations. Prior works~\cite{ShiJXXZWSZY24,MengH0LWWWLM25} showed that Speech-LLMs perform well on multi-talker ASR, but without incorporating speaker diarization. The recent DiarizationLM~\cite{WangHZCXL24} is a post-processing LLM that refines outputs from independent ASR and diarization systems to produce better speaker-attributed transcriptions. However, it is not end-to-end and suffers from error propagation. Concurrently, SpeakerLM~\cite{yin2025speakerlm} trains a Speech-LLM jointly for ASR and diarization in an end-to-end manner, yet it is limited to short audio in local settings and was not compared against recent strong baselines~\cite{abs-2409-06656,WangWDPKMHKBG25,WangHZCXL24}.

In this paper, we propose an end-to-end Speech-LLM for \textbf{J}oint str\textbf{E}amable \textbf{DI}arization and a\textbf{S}r (\textbf{JEDIS-LLM}). Although trained only on short audio ($\le$20s), the model supports chunk-wise streaming inference on arbitrary long-form recordings without additional training. A central challenge of long audio diarization is that splitting audio into short chunks often causes speaker permutation inconsistency~\cite{XueHF0GN21}. Traditional methods address this with post-hoc global clustering~\cite{KinoshitaDT21}. Inspired by prior works~\cite{XueHF0GN21,abs-2507-18446} and the autoregressive property of LLMs, we introduce the \textbf{Speaker Prompt Cache (SPC)} with an on-the-fly update mechanism for chunk-wise streaming inference that preserves speaker consistency across chunks without explicit permutation resolution or retraining. SPC stores a representative utterance for each speaker and combines it with the current chunk during inference. It also naturally supports pre-enrolled \textbf{Speaker Profiles}, which are commonly used in scenarios such as meeting transcription.
During training, we enhance the diarization capability of our proposed model by introducing \textbf{Word-level Speaker Supervision} to the speech encoder, in contrast to prevailing Speech-LLM architectures~\cite{FathullahWLJSLG24,ShiJXXZWSZY24} for ASR that consist only of a speech encoder, a projector, and an LLM.

Experimental results show that our approach outperforms strong baselines, including Sortformer~\cite{abs-2409-06656} and Meta-Cat~\cite{WangWDPKMHKBG25} in the local setting (audio up to 20s), and DiarizationLM~\cite{WangHZCXL24} on long-form audio, while remaining fully end-to-end and streamable, unlike DiarizationLM’s cascaded offline pipeline. To the best of our knowledge, this is the first work to achieve zero-shot, streamable joint ASR and diarization on arbitrary long-form audio using a Speech-LLM trained only on short clips, achieving state-of-the-art results.

\begin{figure*}[t]
    \centering
    \includegraphics[width=\linewidth]{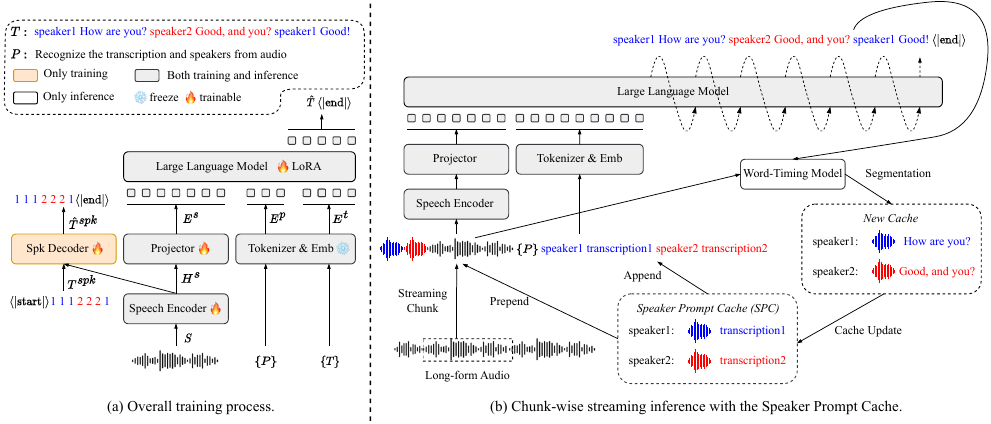}
    \vspace{-0.5cm}
    \caption{Overall training pipeline (a) and chunk-wise streaming inference using the Speaker Prompt Cache (SPC) for long-form audio (b). The Word-Timing Model provides timestamp alignment of each word for segmentation. SPC is not required for offline inference on short audio.}
    \label{fig:1}
    \vspace{-0.3cm}
\end{figure*}

\section{method}

\label{sec:method}
\vspace{-0.2cm}
% \subsection{Backbone: Phi-4-Multimodal}
% \textit{Phi-4-Multimodal} is a state-of-the-art multimodal large language model (LLM) that integrates both audio and visual modalities. At the time of its release, it reached the top rank on the OpenASR leaderboard. In this work, we adopt its audio branch as the backbone. This branch consists of a Speech-Encoder that extracts high-level speech representations, a linear Projector that maps these representations into the text space of the language model, the \textit{Phi-4 Mini} as LLM and a Low-Rank Adaptor (LoRA) specifically designed to optimize the speech modality during training. For convenience, we call it \textbf{Phi-4-MM} in the rest of the paper.

\subsection{Training on Short Audio}
\vspace{-0.1cm}
As shown in Figure~\ref{fig:1} (a), our proposed JEDIS-LLM is built on the prevailing Speech-LLM architecture~\cite{FathullahWLJSLG24,ShiJXXZWSZY24} with speaker-attributed transcription as the LLM objective. The key difference is the addition of a \texttt{Spk-Decoder} for speaker supervision to enhance diarization capability. We describe these components separately below.

\subsubsection{Speech-LLM for joint ASR and Diarization}
\label{sec:train}
\vspace{-0.1cm}
%The training process is illustrated in 
% Figure~\ref{fig:1} (a) illustrates the training process.
We construct speaker-attributed transcriptions by integrating multi-talker transcriptions with speaker IDs, which serve as the LLM training objective. For multi-speaker utterances, words are arranged in temporal order, with a speaker ID inserted whenever the speaker changes, thus forming the \textit{segment-level} objective. In contrast, the \textit{word-level} objective inserts a speaker ID before every word~\cite{abs-2409-06656,WangWDPKMHKBG25}, which under-utilizes the contextual modeling ability of LLMs and slows inference due to longer sequence. Therefore, we adopt the \textit{segment-level} objective in this work.
Given a speech signal $S$, the forward process is defined as:
\vspace{-0.1cm}
\begin{align}
&H^s = \text{Speech-Encoder}(S), \ \
E^s = \text{Projector}(H^s), \\
&E^t = \text{Emb}(\text{Tokenizer}(T)), \ \
E^p = \text{Emb}(\text{Tokenizer}(P)), \\
&\hat{T} = \text{LLM}(\text{Concat}(E^s, E^p, E^t)),
\end{align}
% \vspace{-0.1cm}
where $T$ denotes the speaker-attributed transcription and $P$ is the text prompt. The \texttt{Tokenizer} and \texttt{Emb} correspond to the LLM’s tokenizer and text embedding layers. An additional LoRA~\cite{HuSWALWWC22} is introduced to adapt the LLM outputs to the format required for joint ASR and diarization. Finally, the Cross-Entropy (CE) loss is computed between the predicted sequence $\hat{T}$ and the ground truth $T$:
\vspace{-0.2cm}
\begin{align}
\mathcal{L}_\text{LLM} = \text{CE}(\hat{T}, T).
\end{align}

\vspace{-0.5cm}
\subsubsection{Word-level Speaker Supervision for Speech Encoder}
\vspace{-0.1cm}
% In previous works, two separate encoders were typically employed to encode ASR information and diarization information independently, followed by fusion of the two embeddings. In contrast, we adapt the single Speech Encoder in Speech-LLM jointly encode both types of information.

% To enhance the speaker-awareness capability of the Speech Encoder, we introduce an additional speaker supervision on it. Prior works have commonly employed frame-level multi-class binary classification loss. In contrast, we propose a new scheme termed \textbf{Word-level Speaker Supervision}. As shown in figure~\ref{fig:1}, each word in the transcription is replaced with its corresponding speaker ID, resulting in a word-level speaker ID sequence. This sequence is then predicted by an additional transformer-based \texttt{Spk-Decoder} applied to the output of the Speech Encoder. 
% Finally, the cross-entropy loss is computed between the predicted and reference speaker ID sequences:
% \begin{align}
% &\hat{T}^{spk} = \text{Spk-Decoder}(T^{spk}, H^s), \\
% &\mathcal{L}_\text{Spk} = \text{CE}(\hat{T}^{spk}, T^{spk}).
% \end{align}
% where $T^{spk}$ and $\hat{T}^{spk}$ denote the ground-truth and predicted speaker ID sequences, respectively.

To enhance the speaker diarization capability of the speech encoder, we introduce additional speaker supervision during training to encourage the encoder to learn speaker-discriminative features. While prior works typically adopt \textbf{frame-level} multi-class binary classification loss for speaker diarization~\cite{FujitaKHNW19,abs-2409-06656}, we observe that this approach can harm ASR performance in the joint ASR and diarization task, as frame-level labels lack semantic information. Moreover, such labels are obtained from forced alignment and often contain annotation errors. To address this, we propose a new scheme, termed \textbf{Word-level Speaker Supervision} (distinct from the ``word-level'' objective in Section~\ref{sec:train}). As illustrated in Figure~\ref{fig:1} (a), each word in the transcription is replaced with its corresponding speaker ID, forming a word-level speaker ID sequence. This sequence is predicted by a transformer-based \texttt{Spk-Decoder} applied to the encoder output.
Finally, the CE loss is computed between the predicted and reference sequences:
\vspace{-0.1cm}
\begin{align}
&\hat{T}^{spk} = \text{Spk-Decoder}(T^{spk}, H^s), \ \
\mathcal{L}_\text{Spk} = \text{CE}(\hat{T}^{spk}, T^{spk}),
\end{align}
% \vspace{-0.1cm}
where $T^{spk}$ and $\hat{T}^{spk}$ denote the reference and predicted speaker ID sequences, respectively.

% The speaker supervision module is only employed during training and is omitted at inference time.
% The overall training objective is defined as a weighted sum of the LLM token prediction loss and the speaker supervision loss:
% \begin{align}
% \mathcal{L} = \alpha \cdot \mathcal{L}_\text{LLM} + (1-\alpha) \cdot \mathcal{L}_\text{Spk},
% \end{align}
Since speaker supervision is an auxiliary task aimed at enhancing the encoder’s diarization capability, the \texttt{Spk-Decoder} is used only during training and discarded at inference. The overall objective is a weighted sum of the LLM token prediction loss and the speaker supervision loss:
\vspace{-0.1cm}
\begin{align}
\mathcal{L} = \mu \cdot \mathcal{L}_\text{LLM} + (1-\mu) \cdot \mathcal{L}_\text{Spk},
\end{align}
where $\mu$ is a hyperparameter that balances the two losses.

\begin{table*}[t]
\centering
\caption{Performance comparison of different methods in the local setting (audio up to 20s), reported in WDER (\%) and cpWER (\%). All inference is non-streaming. Phi-4-Multimodal baseline attributes all hypotheses to ``speaker1'' as it does not support speaker diarization.}
\vspace{-0.3cm}
\resizebox{0.88\textwidth}{!}{
\begin{tabular}{c|c|c|cc|cc|cc}
\toprule
\multirow{2}{*}{System} & \multirow{2}{*}{LLM Objective} & \multirow{2}{*}{\shortstack{Speaker Supervision\\for Speech Encoder}} & \multicolumn{2}{c|}{AMI Test} & \multicolumn{2}{c|}{CH109 Full} & \multicolumn{2}{c}{Internal Test Set}  \\
 & & & WDER & cpWER & WDER & cpWER & WDER & cpWER \\
\midrule
% \multicolumn{9}{c}{\textit{Previous works}} \\
% \midrule
Sortformer~\cite{abs-2409-06656} & - & - & - & 26.71 & - & 21.45 & - & - \\
Meta-Cat~\cite{WangWDPKMHKBG25} & - & - & - & 26.02 & - & 26.17 & - & - \\
% \midrule
% \multicolumn{9}{c}{\textit{Phi-based}} \\
% \midrule
Phi-4-Multimodal\footref{phi4-mm}~\cite{abouelenin2025phi} & - & - & 14.52 & 28.09 & 17.25 & 33.09 & 14.68 & 31.10 \\
\midrule
\multirow{3}{*}{JEDIS-LLM (Ablation)} & Segment-level & None & 10.87 & 26.00 & 3.67 & 19.90 & 7.27 & 23.92 \\
 & Segment-level & Frame-level & 8.01 & 35.67 & 2.49 & 25.08 & \textbf{2.44} & 24.34 \\
 & Word-level & Word-level & \textbf{6.34} & 24.08 & 2.40 & 24.55 & 2.65 & 18.77 \\
\midrule
JEDIS-LLM (Final Model) & Segment-level & Word-level & 6.97 & \textbf{23.13} & \textbf{2.06} & \textbf{19.46} & 2.49 & \textbf{18.14} \\
\bottomrule
\end{tabular}}
\label{tab:local}
\vspace{-0.3cm}
\end{table*}

\vspace{-0.2cm}
\subsection{Inference on Long Audio}
\vspace{-0.1cm}
\subsubsection{SPC for Streaming Inference on Long-Form Audio}
\vspace{-0.1cm}
% For long-form audio, performing diarization inference on short chunks may lead to speaker permutation inconsistencies (i.e., the same speaker may be assigned different labels across chunks). Traditional approaches typically employ global clustering as a post-processing step to resolve the global speaker permutation. 

% Here, we propose an alternative method: \textbf{Speaker Prompt Cache (SPC)} for chunk-wise streaming inference, which maintains speaker permutation consistency while eliminating the need for explicit permutation resolution, as shown in Figure~\ref{fig:2}. Specifically, the SPC stores one utterance (both the audio clip and its transcription) for each speaker observed in previous chunks.

% During chunk-wise inference, leveraging the autoregressive property of the LLM, we prepend the cached audio clips to the current chunk and append the cached speaker-attribued transcriptions to the prompt as the initial context for autoregressive generation. The append order follows the speaker indices. This approach enables the model to predict speaker-attribued transcriptions for the current chunk while maintaining speaker consistency.

% Additionally, we design a cache update algorithm to maintain the SPC during inference, the details of streaming inference are shown in 
% Algorithm~\ref{alg:cache_inference}.

For long-form global diarization, inference on short chunks may cause speaker permutation inconsistencies~\cite{XueHF0GN21}, where the same speaker is assigned different labels across chunks. Traditional methods resolve this with global clustering as a post-processing step~\cite{KinoshitaDT21}.

We propose an alternative: the \textbf{Speaker Prompt Cache (SPC)} for chunk-wise streaming inference. As illustrated in Figure~\ref{fig:1} (b), during inference, SPC preserves speaker consistency without explicit permutation resolution by storing one utterance (audio clip and its transcription) for each previously observed speaker.

During chunk-wise inference, leveraging the autoregressive nature of the LLM, we prepend cached audio clips to the current chunk and append cached speaker-attributed transcriptions to the prompt as initial context, ordered by speaker index. This operation provides an exact speaker permutation that serves as the condition for LLM inference, allowing the LLM to follow the same permutation and generate speaker-attributed transcriptions for the current chunk while maintaining speaker consistency.

Additionally, we design a cache update algorithm to maintain the SPC during inference, detailed in Algorithm~\ref{alg:cache_inference}.

\vspace{-0.2cm}
\subsubsection{Seamless Integration with Speaker Profiles}
\vspace{-0.1cm}
% Building upon the idea of the \textbf{Speaker Prompt Cache}, we can seamlessly integrate speaker profiles during inference by replacing the cache with fixed pre-cut audio clips and their corresponding transcriptions, which serve as the speaker profiles. This design offers two key advantages:
% \begin{itemize}
%     \item The exact speaker names can be retrieved from the speaker--ID mapping.
%     \item By fixing the Speaker Prompt Cache with pre-cut high-quality audio clips and transcriptions, the cache no longer requires updates, ensuring more stable performance.
% \end{itemize}

Building on the \textbf{SPC}, we enable seamless integration of \textbf{Speaker Profiles} during inference by replacing the SPC with fixed, manually segmented audio clips and their transcriptions, which serve as speaker profiles. This design provides two main benefits:
\begin{itemize}
    \item The exact speaker names can be retrieved through the speaker-profile mapping. For example, a profile map may be \{speaker1: Mike, speaker2: Susan\}.
    \vspace{-0.1cm}
    \item Using fixed high-quality audio clips and transcriptions instead of on-the-fly SPC eliminates the need for cache updates, yielding more stable performance.
\end{itemize}

\begin{table*}[t]
\centering
\caption{Performance comparison of different approaches in the global setting for long-form audio, reported in terms of WDER (\%) and cpWER (\%). ``Without SPC Update'' refers to not updating the speakers already stored in the Speaker Prompt Cache.}
\vspace{-0.3cm}
\resizebox{\textwidth}{!}{
\begin{tabular}{c|c|c|c|cc|cc}
\toprule
\multirow{2}{*}{System} & \multirow{2}{*}{\shortstack{Strategy to Maintain \\ Global Speaker Consistency}} & \multirow{2}{*}{Streaming Chunks} & \multirow{2}{*}{SPC Update} & \multicolumn{2}{c|}{CH109 Test} & \multicolumn{2}{c}{Fisher Test}  \\
 & & & & WDER & cpWER & WDER & cpWER \\
\midrule
\multicolumn{8}{l}{\textit{Non-streaming Inference}} \\
\midrule
DiarizationLM (Llama 3)~\cite{WangHZCXL24} & \multirow{2}{*}{\shortstack{Independent ASR \& Diarization \\ with LLM Post Processing}} & - & - & 6.66 & 23.57 & 3.28 & 18.37 \\
DiarizationLM (PaLM 2)~\cite{WangHZCXL24} & & - & - & 4.25 & 20.22 & 2.37 & 16.93 \\
\midrule
JEDIS-LLM & Offline Chunk Inference + Global Clustering & - & - & 2.48 & 19.03 & 2.06 & \textbf{15.03} \\
\midrule
\multicolumn{8}{l}{\textit{Chunk-wise Streaming Inference}} \\
\midrule
\multirow{4}{*}{JEDIS-LLM} & \multirow{4}{*}{\shortstack{Streaming Inference with \\ Speaker Prompt Cache (SPC)}} & \multirow{2}{*}{Oracle Chunks} & \xmark & 2.09 & 18.58 & 2.51 & 16.40 \\
 & & & \cmark & \textbf{1.73} & \textbf{18.20} & \textbf{2.05} & 15.88 \\
\cline{3-8}
 & & \multirow{2}{*}{VAD Chunks} & \xmark & 2.62 & 19.32 & 2.95 & 17.37 \\
 & & & \cmark & 2.54 & 19.09 & 2.35 & 16.60 \\
\bottomrule
\end{tabular}}
\label{tab:global}
\vspace{-0.3cm}
\end{table*}

\begin{table}[t]
\centering
\caption{Comparison of streaming inference with and without speaker profiles on the CH109 test set for long-form audio in the global setting, showing cpWER (\%), SA-WER (\%), and their difference ($\Delta$), where $\Delta$ indicates how strictly predicted speaker IDs match the reference.}
\vspace{-0.3cm}
\resizebox{\columnwidth}{!}{
\begin{tabular}{c|c|ccc}
\toprule
Streaming Chunks & Speaker Profiles & cpWER & SA-WER & $\Delta$ \\
\midrule
\multirow{2}{*}{Oracle Chunks} & \xmark & 18.20 & 25.98 & 7.78 \\
 & \cmark & 17.91 & 19.98 & 2.07 \\ \hline
\multirow{2}{*}{VAD Chunks} & \xmark & 19.09 & 30.79 & 11.7 \\
 & \cmark & 19.18 & 21.94 & 2.76 \\
\bottomrule
\end{tabular}}
\label{tab:profile}
\end{table}

\vspace{-0.2cm}
\section{Experimental Settings}
\label{sec:exp_set}
\vspace{-0.1cm}

\subsection{Training Datasets}
\vspace{-0.1cm}

We trained the proposed JEDIS-LLM on five data sources. Public datasets include the AMI Corpus~\cite{CarlettaABFGHKKKKLLLMPRW05} (train and dev sets, IHM-Mix channel, up to 4 speakers/session, $\sim$90~h), the ICSI Corpus~\cite{JaninBEEGMPPSSW03} (IHM-Mix channel, all subsets, up to 11 speakers/session, $\sim$71~h), and the Fisher Corpus~\cite{cieri2004fisher} (2 speakers/session, $\sim$1929~h). We also incorporated internally collected data (up to 7 speakers/session, $\sim$6734~h) and simulated conversations from VoxCeleb1~\cite{NagraniCXZ20} and VoxCeleb2~\cite{ChungNZ18} ($\sim$964~h). For the simulations, we removed non-English utterances using language identification~\cite{speechbrain,valk2021slt}, and mixed 5 speakers per conversation, each contributing 3–5 sentences with mild overlap ($\leq$1\%) and room impulse responses up to 0.2~s. In total, the training data amounts to about 10k~hours.

\vspace{-0.3cm}
\subsection{Training Setting}
\vspace{-0.1cm}
% Our model is built upon Phi-4-Multimodal~\cite{abouelenin2025phi}, a state-of-the-art multimodal LLM. We use it's speech branch as the initialize of our system. An additional LoRA module is applied for adapting to speaker-attributed transcription, configured with an $\alpha$ of 32 and a rank of 16. The \texttt{Spk-Decoder} for speaker-aware supervision consists of 3 transformer decoder layers, each with 1024-dimensional hidden states, 16 attention heads, and 1024-dimensional feed-forward layers. For long-form audio, we randomly segment samples into 15--20 second clips during training. The model is trained on 16 NVIDIA A100 80GB GPUs with an effective batch size of 256 seconds per GPU. We employ DeepSpeed Stage-1 for distributed training. The AdamW optimizer is used with a peak learning rate of 0.0001. A linear warmup-decay learning rate scheduler is applied, with 1000 warmup steps and a maximum of 40{,}000 training steps.

Our model builds on Phi-4-Multimodal\footnote{\url{https://huggingface.co/microsoft/Phi-4-multimodal-instruct}\label{phi4-mm}}~\cite{abouelenin2025phi}, using its speech branch as initialization of speech encoder, projector and LLM. The additional LoRA is configured with $\alpha=32$ and $rank=16$. The \texttt{Spk-Decoder} has 3 transformer layers with 1024-dimensional hidden states, 16 attention heads, and 1024-dimensional feed-forward layers. The loss weight $\mu$ is set to 0.5. Long-form audio is randomly segmented into 15$\sim$20s clips for training. The model is trained on 16 NVIDIA A100 80GB GPUs with 256s per GPU batch size, using AdamW optimizer (peak learning rate 0.0001) with linear warmup-decay scheduling (1000 warmup steps, 40{,}000 steps total).

\begin{algorithm}[t]
\fontsize{8pt}{9pt}\selectfont
% \footnotesize
\caption{\footnotesize Streaming Inference with Speaker Prompt Cache}
\label{alg:cache_inference}

\SetAlCapSkip{0pt}                  % 算法标题与内容间距
\SetInd{0.55em}{0.55em}               % {一级缩进}{子缩进}，默认为1em
\SetAlgoInsideSkip{0pt}             % 算法环境内部上下间距

\SetKwFor{ForEach}{for each}{do}{end}
\KwIn{Long audio $A$, well-trained model $M$, $prompt$}
\KwOut{Speaker-attributed transcriptions $R$}
Initialize empty speaker prompt cache $C$, result list $R$ \\
Define profile audio length threshold $l$, text length threshold $n$\;
Define dvector similarity threshold $\theta$\;
\ForEach{chunk $a$ in $A$}{
    % \tcp{Inference with Cache}
    \eIf{$C=\emptyset$}{
        $a_{\text{infer}}\gets a,\; p_{\text{infer}}\gets prompt$\;
    }{
        $a_{\text{infer}}\gets\{C[s].audio,\forall s\in C\}+a$\;
        $p_{\text{infer}}\gets prompt+\{C[s].text,\forall s\in C\}$\;
    }
    $r\gets M(a_{\text{infer}},p_{\text{infer}})$, Append $r$ to $R$\;

    % \tcp{Cache update}
    \ForEach{speaker $s$ in $r$}{
        $Ali \gets \text{WordTimingModel}(a, s.text)$\;
        $(A_s, T_s) \gets \text{Segmentation}(Ali, \text{len} \leq l, \text{exclude overlap})$\;
        $(\hat{a}_s,\hat{t}_s)\gets$ FindLongest$(A_s,T_s)$\;

        \If{$s\notin C$}{
            $C[s].audio\gets\hat{a}_s,\; C[s].text\gets\hat{t}_s$\;
        }
        \ElseIf{len$(C[s].text)<n$ \textbf{or} $\neg$HasPunctuation$(C[s].text)$}{
            \If{len$(\hat{a}_s)>$len$(C[s].audio)$}{
                $\sigma\gets$ CosSim(dvector$(\hat{a}_s)$, dvector$(C[s].audio))$\;
                \If{$\sigma>\theta$}{
                    $C[s].audio\gets\hat{a}_s,\; C[s].text\gets\hat{t}_s$\;
                }
            }
        }
    }
}
\Return $R$\;
\end{algorithm}

\vspace{-0.3cm}

\subsection{Evaluation Setting}
\vspace{-0.1cm}

We evaluate our JEDIS-LLM under both \textbf{local} (short audio) and \textbf{global} (long audio) settings. In the local setting, following prior works~\cite{abs-2409-06656,WangWDPKMHKBG25}, we evaluate on two datasets: the AMI-IHM-Mix test set (\emph{AMI Test}; up to 4 speakers per session) and the 2-speaker subset of 109 sessions from the Callhome American English Speech corpus (\emph{CH109 Full})~\cite{LDC97S42}. In both datasets, long-form audio is segmented into 10$\sim$20s clips. We additionally evaluate on an internal test set (with up to 5 speakers per session).

For the global setting, following prior work~\cite{WangHZCXL24}, we use test subset of CH109 and Fisher, referred to as \emph{CH109 Test} and \emph{Fisher Test}. In chunk-wise streaming inference, we consider \textbf{Oracle Chunks}, derived from ground-truth sentence boundaries, and \textbf{VAD Chunks}, obtained via voice activity detection\footnote{\url{https://github.com/snakers4/silero-vad}}. Chunks are up to 10 seconds long, and regions without transcriptions at the beginning and end of the long audio are excluded. The dvector extractor in Algorithm~\ref{alg:cache_inference} is a well-trained Res2Net~\cite{ZhouZW21} trained for speaker verification, and the Word-Timing Model is an internal forced alignment model. The profile audio length threshold $l$ and text length threshold $n$ are set to 5 seconds and 8, respectively. The dvector similarity threshold $\theta$ is set to 0.7.
For speaker profile integration, we evaluate on \emph{CH109 Test} by extracting audio clips shorter than 5 seconds from non-transcribed portions and manually annotating them as speaker profiles.
% In the global setting (streaming inference and profile integration), we evaluate the model without CoT reasoning.

We report \textbf{Word Diarization Error Rate (WDER)}~\cite{ShafeySS19} and \textbf{concatenated minimum-permutation Word Error Rate (cpWER)}~\cite{watanabe2020chime} as primary metrics. Pure WER is ambiguous in multi-speaker scenarios and is thus excluded. For speaker profiles, we also report \textbf{Speaker-Attributed WER (SA-WER)}~\cite{KandaGWMCZY20}, which directly matches predicted speaker IDs with the reference without finding best permutation, providing a stricter measure than cpWER.

\vspace{-0.2cm}
\section{Experimental Results}
\vspace{-0.1cm}
\subsection{Evaluation on Local Setting for Short Audio}
\vspace{-0.1cm}
Table~\ref{tab:local} presents the performance comparison in the local setting (audio clips under 20 seconds) using non-streaming inference. Compared with previous strong baselines (Sortformer~\cite{abs-2409-06656} and Meta-Cat~\cite{WangWDPKMHKBG25}), our JEDIS-LLM achieves significantly better cpWER, demonstrating both the advantages of Speech-LLMs for joint ASR and diarization and the effectiveness of our proposed method.

The ablation study further shows that speaker supervision on the encoder is crucial for diarization, as removing it leads to higher WDER. Frame-level speaker supervision~\cite{FujitaKHNW19,abs-2409-06656} improves diarization performance compared to no speaker supervision, but degrades cpWER, indicating that frame-level loss negatively affects the encoder’s ASR capability. Using word-level speaker-attributed transcription as the LLM objective yields reasonable WDER performance, particularly on AMI where speaker turns are frequent, but its cpWER is still worse than that of segment-level transcription. This is because word-level speaker-attributed transcription introduces excessive speaker label splits, disrupting context. Overall, combining segment-level transcription for the LLM with word-level speaker supervision for the encoder yields the best performance.

\vspace{-0.3cm}
\subsection{Evaluation on Global Setting for Long-form Audio}
\vspace{-0.1cm}
Table~\ref{tab:global} presents the performance comparison in the global setting for long-form audio. Previous work, DiarizationLM~\cite{WangHZCXL24}, aligns independent ASR and speaker diarization results and subsequently employs a finetuned LLM for post-processing. We evaluate our model under both offline and chunk-wise streaming inference. For offline inference, long-form audio is segmented into chunks under 20 seconds based on oracle sentence boundaries, followed by offline inference. The word-timing model and global clustering are then applied to produce global results. For chunk-wise streaming inference, we employ the Speaker Prompt Cache (SPC) to maintain speaker consistency across chunks.

The results show that our proposed streaming JEDIS-LLM produces substantially better performance, significantly outperforming DiarizationLM using either oracle or VAD chunks, despite the latter being a cascaded system using offline post-processing. As expected, enabling the SPC update mechanism improves performance by refreshing the cache with higher-quality speaker prompts. Moreover, streaming inference surpasses the ``Offline Chunk Inference + Global Clustering'' baseline on the CH109 test set and achieves the best WDER on the Fisher test set when using oracle chunks. These results demonstrate the effectiveness of SPC and its update strategy. 
% We consider this a significant breakthrough, as it enables streaming inference for joint ASR and diarization on long audio using a Speech-LLM trained solely on short audio, while achieving state-of-the-art performance.

\vspace{-0.3cm}
\subsection{Performance of Speaker Profiles Integration}
\vspace{-0.1cm}
% Table~\ref{tab:profile} reports the results of streaming Phi-4-SD with and without speaker profiles in terms of cpWER (\%), SA-WER (\%), and their difference ($\Delta$). In the setting without profiles, reference speaker IDs are assigned according to the order of arrival (speaker1, speaker2, \ldots), while in the setting with profiles, they follow the order in which the profiles are concatenated. The results show that incorporating speaker profiles leads to a smaller gap between SA-WER and cpWER, suggesting that using manually defined, fixed, high-quality speaker profiles rather than an on-the-fly updated SPC helps the predicted speaker IDs better align with the reference speakers. Moreover, providing speaker profiles allows the system to directly map predicted speaker IDs to actual speaker names by speaker--profile mapping, rather than index-type labels, which is particularly beneficial for real-world applications.

Table~\ref{tab:profile} presents the results of streaming inference with and without speaker profiles on long-form audio in the global setting, evaluated by cpWER (\%), SA-WER (\%), and their difference ($\Delta$). Without profiles, reference speaker IDs are assigned by order of appearance (speaker1, speaker2, \ldots), whereas with profiles they follow the concatenation order of the given profiles. The results show that incorporating profiles narrows the gap between SA-WER and cpWER, indicating that fixed, manually segmented profiles align predicted IDs with reference speakers more effectively than an on-the-fly automatically updated SPC. In addition, profiles allow direct mapping from predicted IDs to real speaker names, rather than index labels, which is especially valuable for real-world applications.

\vspace{-0.2cm}
\section{Conclusion}
\label{sec:conclusion}
\vspace{-0.2cm}
In this work, we propose an end-to-end Speech-LLM for joint ASR and speaker diarization, trained only on short audio segments under 20 seconds, yet capable of performing chunk-wise streaming inference on long-form audio without additional training. We introduce a speaker prompt cache with an on-the-fly update mechanism, which enables chunk-wise streaming inference while preserving speaker consistency across chunks. Furthermore, replacing the speaker prompt cache with manually defined high-quality utterances allows seamless integration with speaker profiles. In addition, incorporating word-level speaker supervision into the speech encoder during training enhances the model’s diarization capability. Experimental results show that our approach outperforms strong baselines, including Sortformer and Meta-Cat in the local diarization setting (up to 20 seconds), and DiarizationLM in the global setting for long-form audio, while remaining fully end-to-end and streamable, in contrast to DiarizationLM’s cascaded offline pipeline. To the best of our knowledge, this is the first work to enable streamable joint ASR and diarization on long audio using a Speech-LLM trained only on short audio, achieving state-of-the-art performance.

% In addition, we investigated the role of chain-of-thought reasoning, and observed that the predicted number of speakers followed the reasoning chain.

% For long-form audio in the global setting, we introduce a speaker prompt cache with an update mechanism, enabling chunk-wise streaming inference. Experiments show that streaming Phi-4-SD significantly outperforms DiarizationLM, an offline post-processing LLM applied to ASR and diarization outputs.

% Furthermore, replacing the prompt cache with fixed high-quality audio clips and transcripts allows seamless integration with speaker profiles. This leads to outputs speaker IDs better align with reference speakers during inference. 

% For long-form audio in the global setting, we introduce a speaker prompt cache with an update mechanism, enabling chunk-wise streaming inference. Experiments show that streaming Phi-4-SD significantly outperforms DiarizationLM, an offline post-processing LLM applied to ASR and diarization outputs. Moreover, replacing the speaker prompt cache with pre-defined fixed high-quality audio clips and transcripts enables seamless integration with speaker profiles, resulting in output speaker IDs more accurately aligned with reference speakers. In conclusion, this work represents the first unified LLM-based framework for joint ASR and speaker diarization in both streaming and non-streaming scenarios, advancing the state of the art and marking a significant advancement in the field.

\vfill\pagebreak

\newpage

\newpage
\newpage

\section{appendix}
\subsection{Detail of Training Data}
The statistics of the training dataset are shown in Table~\ref{tab:train_data}.

\vspace{0.2cm}
\begin{table}[h]
\centering
\caption{Statistics of the training data sets.}
\vspace{-0.2cm}
\resizebox{\columnwidth}{!}{
\begin{tabular}{c|c|c|c}
% \cline{1-4}
\toprule
Dataset        & \#sessions & \#speakers per session & Duration (hours) \\
\midrule
AMI train\&dev & 155        & 4          & 90 \\
% \midrule
ICSI all       & 75         & 3$\sim$11  & 71 \\
% \midrule
Fisher train   & 11,527     & 2          & 1929 \\
% \midrule
% Call center    & 30,794     & 2          & 2864 \\
% % \midrule
% Podcast        & 5,324      & 2$\sim$7   & 3870 \\
Internal    & 36,118 & 2$\sim$7 & 6734
\\
% \midrule
Simulated      & 21,943     & 5          & 964 \\
\bottomrule
\end{tabular}}
\label{tab:train_data}
\end{table}

\subsection{Different Chunk and SPC Lengths for Streaming Inference}
Beyond the default experimental setting, we further explore different chunk lengths and per-speaker SPC lengths in streaming inference for long-form audio. As shown in Table~\ref{tab:chunk}, reducing the SPC length to 3 seconds proves beneficial for ASR. This is because the text in SPC is a hypothesis rather than ground truth; shorter SPCs contain fewer errors, which improves the quality of the initial context for LLM inference. In contrast, reducing the chunk length degrades overall performance, since shorter chunks make it more difficult to find high-quality segments within the chunk for updating the SPC on-the-fly.
\vspace{0.3cm}
\begin{table}[h]
\centering
\caption{Performance comparison of different chunk lengths and per-speaker SPC lengths in streaming inference for long-form audio, reported in terms of WDER (\%) and cpWER (\%).}
\vspace{-0.2cm}
\resizebox{\columnwidth}{!}{
\begin{tabular}{c|c|c|cc|cc}
\toprule
\multirow{2}{*}{\shortstack{Streaming\\Chunks}} & \multirow{2}{*}{\shortstack{Chunk\\Length}} & \multirow{2}{*}{\shortstack{SPC\\Length}} & \multicolumn{2}{c|}{CH109 Test} & \multicolumn{2}{c}{Fisher Test}  \\
 & & & WDER & cpWER & WDER & cpWER \\
\midrule
\multirow{2}{*}{\shortstack{Oracle\\Chunks}} & $\le10s$ & $\le5s$ & 1.73 & 18.20 & 2.05 & 15.88 \\
 & $\le10s$ & $\le3s$ & 2.11 & 18.43 & 2.05 & 15.65 \\
\midrule
\multirow{5}{*}{\shortstack{VAD\\Chunks}} & $\le10s$ & $\le5s$ & 2.54 & 19.09 & 2.35 & 16.60 \\
 & $\le10s$ & $\le3s$ & 2.55 & 18.91 & 2.20 & 16.27 \\
 & $\le5s$ & $\le5s$ & 2.85 & 23.38 & 2.91 & 18.35 \\
 & $\le5s$ & $\le3s$ & 2.88 & 21.46 & 2.93 & 18.18 \\
 & $\le3s$ & $\le3s$ & 4.29 & 25.54 & 3.83 & 21.03 \\
\bottomrule
\end{tabular}}
\label{tab:chunk}
\vspace{-0.3cm}
\end{table}

\subsection{Chain-of-Thought Exploration}
% Since joint ASR and diarization in multi-talker scenarios is challenging, 
% we explore the use of Chain-of-Thought (CoT) reasoning within the speech-LLM. 
% Specifically, we prepend a simple reasoning chain before the transcription to first estimate the number of speakers in the audio clip, e.g., \textit{	``estimated\_speaker\_number: 3''}. 
% We denote this reasoning chain as $C$, and its text embedding as $E^c$. 
% Accordingly, Eq.~(5) can be modified as:
% \begin{align}
% \text{Concat}(\hat{C}, \hat{T}) = \text{LLM}(\text{Concat}(E^s, E^p, E^c, E^t)),
% \end{align}
% where the token prediction loss is computed jointly over both the reasoning chain and the transcription. 
% This encourages the model to first estimate the number of speakers through the reasoning chain, thereby improving subsequent diarization.

Since joint ASR and diarization in multi-talker scenarios is challenging, 
we explore the use of Chain-of-Thought (CoT) reasoning within the Speech-LLM. 
Specifically, we prepend a simple reasoning chain before the speaker-attributed transcription to estimate the number of speakers in the input audio, e.g., \textit{``estimated\_speaker\_number: 3''}. 
We denote the reasoning chain as $C$ with embedding $E^c$, and modify Eq.~(5) as:
\begin{align}
\text{Concat}(\hat{C}, \hat{T}) = \text{LLM}(\text{Concat}(E^s, E^p, E^c, E^t)),
\end{align}
where the token prediction loss is computed jointly over the reasoning chain and the transcription. 
% This encourages the model to explicitly infer the speaker number before transcription, thereby enhancing diarization performance.

The results on the AMI test set are shown in Table~\ref{tab:cot}. We observe that incorporating CoT yields slightly better WDER and speaker counting accuracy, particularly for utterances involving four speakers, suggesting that CoT helps in estimating larger speaker numbers. When using ground-truth CoT that includes the true number of speakers for inference, WDER is further reduced and speaker counting accuracy approaches 100\%, indicating that the reasoning chain can effectively guide speaker-attributed transcription. 
% This approach is practically valuable, since in real-world scenarios the number of speakers in a meeting is often known in advance.
However, even when the number of speakers was estimated accurately, the cpWER did not improve, and the WDER improvement was also limited. This suggests that the model did not truly capture the conversational dynamics when provided with the actual number of speakers in the reasoning chain.

\vspace{0.3cm}
\begin{table}[h]
\centering
\caption{Performance comparison of our JEDIS-LLM with and without Chain-of-Thought (CoT) on the AMI test set, reported in terms of WDER (\%), cpWER (\%), and Speaker Counting Accuracy (\%). For speaker counting, the number of speakers is taken from the predicted speaker-attributed transcriptions, rather than from the reasoning chain.}
\vspace{-0.2cm}
\resizebox{\columnwidth}{!}{
\begin{tabular}{c|cc|ccccc}
\toprule
\multirow{2}{*}{CoT Type} & \multirow{2}{*}{WDER} & \multirow{2}{*}{cpWER} & \multicolumn{5}{c}{Speaker Counting Accuracy} \\
 & & & 1-spk & 2-spk & 3-spk & 4-spk & avg \\
\midrule
w/o CoT & 6.97 & \textbf{23.13} & 96.1 & 84.4 & 60.0 & 30.4 & 68.9 \\
Predicted & 6.93 & 23.43 & 93.7 & 82.2 & 61.0 & 37.0 & 69.5 \\
Ground-truth & \textbf{6.60} & 23.83 & \textbf{100} & \textbf{100} & \textbf{98.5} & \textbf{98.2} & \textbf{99.2} \\
\bottomrule
\end{tabular}}
\label{tab:cot}
% \vspace{0.3cm}
\end{table}

\end{document}